\documentclass{aa}
\usepackage{psfig}
\newcommand{\apj}[2]{ApJ #1, #2}

\newcommand{\aaa}[2]{A\&A #1, #2}

\newcommand{\mon}[2]{MNRAS #1, #2}

\newcommand{\cts}{counts s$^{-1}$}

\newcommand{\rxja}{RX\,J1856.6--3754}
\newcommand{\rxjb}{RX\,J0720.4--3125}
\newcommand{\rxjc}{RX\,J0806.4--4123}
\newcommand{\rxjd}{1RXS\,J130848.6+212708}
\newcommand{\rxje}{RX\,J1605.3+3249}
\newcommand{\rxjf}{RX\,J0420.0--5022}

\newcommand{\tento}[1]{$\times 10^{#1}$}
\newcommand{\hcm}[1]{$\times 10^{#1}$ cm$^{-2}$}
\newcommand{\ergcm}[1]{$\times 10^{#1}$ erg cm$^{-2}$ s$^{-1}$}

\newcommand{\et}{et al.}
\newcommand{\fx}{f$_{\rm x}$}
\newcommand{\fo}{f$_{\rm opt}$}
\begin{document}

\thesaurus{06; (08.09.2: \rxjf; 
                08.14.1;        
                13.25.5)        
           }

\title{\rxjf: An isolated neutron star candidate with evidence for
22.7 s X-ray pulsations$^{\ast}$}

\author{F.\,Haberl\inst{1}, W.\,Pietsch\inst{1} and C.\,Motch\inst{2}}
\offprints{F.\,Haberl (fwh@mpe.mpg.de)\\
           $^{\ast}$ Partly based on NTT observations performed at the
           European Southern Observatory, La Silla, Chile}

\institute{$^1$Max-Planck-Institut f\"ur extraterrestrische Physik,
               Giessenbachstra{\ss}e, 85748 Garching, Germany\\
           $^2$Observatoire de Strasbourg, 11, rue de l'Universite,
               67000 Strasbourg, France\\
          }
\date{Received date; accepted date}

\maketitle\markboth{F. Haberl \et: \rxjf: An isolated neutron star candidate}
                   {F. Haberl \et: \rxjf: An isolated neutron star candidate}
\begin{abstract}
We report the discovery of a new isolated neutron star candidate, \rxjf,
showing evidence ($4\sigma$) for 22.7 s X-ray pulsations in ROSAT data. 
NTT observations of the field around the soft X-ray source do not reveal
any likely optical counterpart brighter than B = 25.25 implying an X-ray to
optical flux ratio of $>10^{3.3}$ and ruling out other possible kinds of X-ray 
emitters. The X-ray spectrum of \rxjf\ 
can be described with blackbody emission with temperature kT of $\sim$57 eV
and four ROSAT detections are consistent with no flux variations on time scales
of years. The X-ray pulsations, if confirmed, make \rxjf\ the second
long-period isolated neutron star candidate after \rxjb. As for this latter source
similar conclusions about the magnetic field strength of the neutron star can be
drawn depending on evolutionary scenarios.

\end{abstract}

\section{Introduction}
To date five candidates for isolated neutron stars (INS) were discovered
in ROSAT data (\rxja, Walter \et\ 1996; \rxjb, Haberl \et\ 1997;
\rxjc, Haberl \et\ 1998; \rxjd, Schwope \et\ 1999 and \rxje, Motch
\et\ 1999) which can be classified by their similar X-ray properties.
The soft X-ray spectra are well represented by pure blackbody emission
with temperatures kT between 50 and 120 eV and the X-ray flux is constant
on time scales of months to 10 years. The X-ray brightest source, \rxja,
could be identified with a faint blue object (V = 25.6) by Walter \&
Matthews (1997) on HST images and for the second brightest, \rxjb, a likely
counterpart with B = 26.1-26.6 was found (Motch \& Haberl 1998,
Kulkarni \& van Kerkwijk 1998). The inferred extremely
high X-ray to optical flux ratios log(\fx/\fo) of 4.8 and 5.3,
respectively, exclude any known object other than an isolated neutron
star. Similarly high lower limits for \fx/\fo\ are derived from 
the faint brightness limits obtained for the other candidates 
(see table in Schwope \et\ 1999).

The five candidates constitute the bright end of the log N -- log S
distribution of X-ray detected INS (with no detectable radio mission)
of which up to several thousand were expected to be seen in the ROSAT
all-sky survey (RASS, Blaes \& Madau 1993, Colpi \et\ 1993, Madau \&
Blaes 1994). The estimates were based on models of old neutron stars
heated by the accretion of interstellar matter as proposed by Ostriker
\et\ (1970). From a compilation of available
data on the number and space density of INS candidates and upper limits
obtained from RASS follow-up identification programs Neuh\"auser
\& Tr\"umper (1999) concluded that the log N -- log S curve lies between
the theoretical expectations for middle-aged cooling neutron stars and old
accreting neutron stars. They suggest that the larger number of expected
accreting old neutron stars than observed is mainly caused by the 
assumed velocity distribution of the neutron stars.

From one of the five INS candidates X-ray pulsations were detected,
indicating the spin period of the neutron star. The pulse period of
8.391 s found for \rxjb\ constrains the magnetic field strength
of the neutron star. In the case of an accreting neutron star
a weak field of less than 10$^{10}$ G is derived for accretion to
be possible (Haberl \et\ 1997). A higher magnetic field strength in the
past is required to spin down the neutron star to the present rate within
a Hubble time {\it if} the neutron star was born with a typical spin period of
10 ms. Consequently this would indicate magnetic field decay and for a
birth field strength of 10$^{12}$ G decay time scales of $>10^7$ y are
derived (Wang 1997) for an old ($>10^9$ y) neutron star. 
Alternatively \rxjb\ may prove to be a relatively
young ($<10^6$ y) cooling neutron
star for which the argumentation from above would require a very high
magnetic field strength of the order of 10$^{14}$ G (Kulkarni \&
Kerkwijk 1998). The population of these "magnetars" was suggested to
explain the properties of soft $\gamma$-ray repeaters (Kouveliotou \et\
1999 and references therein). The extremely strong magnetic field can
suppress radio emission and supply a significant
source of heat, allowing magnetars to remain detectable in X-rays over
longer times than ordinary pulsars (Heyl \& Kulkarni 1998).
However a doubling of the pulse period derivative observed in SGR 1900+14
casts doubt on the magnetar model as it requires a 100\% increase
of the magnetic field energy if magnetic dipole radiation were the primary
cause of the pulsar spin-down (Marsden \et\ 1999). Also the 
different X-ray spectrum of \rxjb\ in comparison to those of soft 
$\gamma$-ray repeaters would need to be explained by the magnetar model.

As long as the distribution of initial spin period / magnetic field strength
is highly uncertain, conclusions on the evolution of these important 
neutron star parameters are tentative.
The nature of \rxjb\ remains therefore unclear until an accurate
measurement of the period derivative may further constrain the models.

In this letter we report on ROSAT X-ray and NTT optical observations
of a new INS candidate, \rxjf, which shows evidence for 22.7 s
X-ray pulsations. Together with \rxjb\ it is only the second long-period
INS candidate and their investigation is crucial for our understanding
of their evolutionary status within the whole class of isolated neutron 
stars.

\section{Observations}

\subsection{Soft X-rays}

\rxjf\ was discovered as X-ray source in the ROSAT all-sky survey
data and originally associated with the galaxy ESO\,202-G008. It is
included in the ROSAT all-sky survey bright source catalogue
(1RXS\,J042003.1--502300, Voges \et\ 1996) with 0.12$\pm$0.03 \cts.
Details on the ROSAT mission can be found in Tr\"umper (1982) and the
focal plane instruments PSPC (Position Sensitive Proportional
Counter) and HRI (High Resolution Imager), both sensitive in the 
0.1 -- 2.4 keV energy range, are described by Pfeffermann \et\ (1986)
and David \et\ (1993), respectively. From two follow-up HRI
observations in June and December 1997 (the observations are summarized 
in Tab.~\ref{'tabobs'}) it became clear that the soft X-ray source is 
nearly 1\arcmin\ away from ESO\,202-G008 and unrelated.
The best position was derived from the June 1997 HRI observation using the 
maximum likelihood technique of EXSAS (Zimmermann \et\ 1994) to
RA = 04$^{\rm h}$ 20$^{\rm m}$ 2\fs2, Dec = \hbox{-50\degr} 22\arcmin\ 46\arcsec\ (J2000.0)
with a 90\% confidence error of 8\arcsec\ (dominated by the 7\arcsec\
systematic error of the attitude reconstruction).
No other sources were detected in the HRI image which could be utilized
for bore-sight correction.
\begin{table}
\caption[]{ROSAT soft X-ray detections of \rxjf\ (0.1 -- 2.4 keV)}
\begin{flushleft}
  \begin{tabular}{lrc}
  \hline
  \noalign{\smallskip}
  Observation                      & exposure & observed         \\
                                   & s~~~~    & counts s$^{-1}$  \\
  \noalign{\smallskip}\hline\noalign{\smallskip}
  Survey \hfill Jan. 23-25, 1991   &      193 & 0.12  $\pm0.03$  \\
  PSPC   \hfill Feb. 20, 1997      &     1538 & 0.11  $\pm0.01$  \\
  HRI    \hfill June 27-29, 1997   &     5782 & 0.021 $\pm0.002$ \\
  HRI    \hfill Dec. 15, 1997      &     1163 & 0.016 $\pm0.004$ \\
  \noalign{\smallskip}
  \hline
  \end{tabular}
  \end{flushleft}
  \label{'tabobs'}
\end{table}

\begin{figure}[]
\psfig{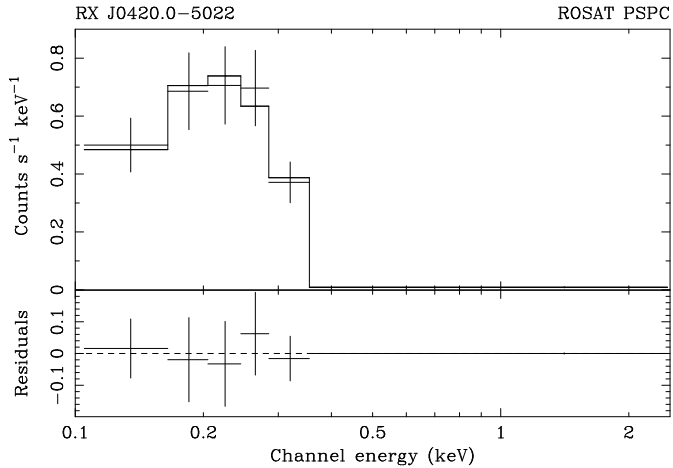}
\caption[]{The PSPC spectrum of \rxjf\ with the best fit blackbody model.
           The residuals are shown in the lower panel}
  \label{'pspcsp'}
\end{figure}

\rxjf\ was serendipitously observed in a short PSPC pointing
in February 1997 at an off-axis angle of 31\arcmin. The count rate 
is consistent with the survey detection and with the
HRI detections in June and December 1997 (see Tab.~\ref{'tabobs'} and 
below).
The average PSPC spectrum (Fig.~\ref{'pspcsp'}), although of low
statistical quality, is very similar to the spectra of the known INS
candidates without any significant number of counts above 0.4 keV. 
A power-law fit to the spectrum is acceptable but results in an 
unrealistic photon index of $\sim$10.
The soft emission is well described by a blackbody model
(kT = 57$^{+25}_{-47}$ eV, column density 1.7\hcm{20} (0 -- 8\hcm{20}))
with an observed flux of 6.9\ergcm{-13} (0.1 -- 2.4 keV).
Assuming a distance d for the source a bolometric luminosity of 
L$_{\rm bol}$ = 2.7\tento{30} (d/100 pc)$^2$ erg s$^{-1}$ is derived. 
Assuming an emission area with 10 km radius, the derived upper limit for the 
blackbody temperature (104 eV) yields un upper limit for the 
distance of 3.9 kpc.
For the best fit temperature the distance is 700 pc but is further 
reduced for a smaller emitting area. Folding the
blackbody model with the best fit parameters through the HRI energy
response yields for the observed PSPC count rate (pointing) an expected
HRI count rate of 1.9\tento{-2} \cts. The PSPC position of 
RA = 04$^{\rm h}$ 20$^{\rm m}$ 1\fs6, Dec = \hbox{-50\degr} 22\arcmin\ 44\arcsec\ 
(J2000.0, error of 17\arcsec) is 6.1\arcsec\ away from the HRI position.
The survey position is 16.5\arcsec\ from the HRI position, also 
consistent within the errors. A formal average of the positions 
(weighting with errors) yields RA = 04$^{\rm h}$ 20$^{\rm m}$ 2\fs36, Dec = 
\hbox{-50\degr} 22\arcmin\ 49\farcs5 (7\farcs4 error), somewhat south 
of the HRI position.

\begin{figure}[]
\psfig{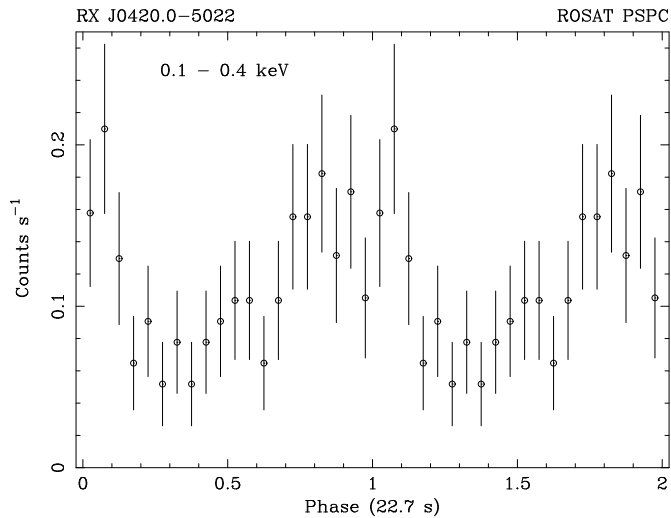}
\caption[]{PSPC light curve from February 20, 1997, folded with a period of 
  22.7 s, for clarity data points are repeated for a second period}
  \label{'pspcflc'}
\end{figure}

A temporal analysis of the PSPC data using Fourier transformation
and a Rayleigh Z$^2$ folding test (Buccheri \et\ 1983) reveals evidence 
for periodic modulation of the soft X-ray flux with 22.69$\pm$0.03 s. 
The probability for a false detection of the period was
derived to 7\tento{-5} corresponding to a 4 $\sigma$ detection.
The PSPC light curve folded by the pulse period is plotted in 
Fig.~\ref{'pspcflc'} and shows a sinusoidal semi-amplitude modulation of
43$\pm$14\%. The folding analysis was also applied to the June 1997 HRI 
observation. The periodogram shows peaks around the expected period, 
supporting the reality of the modulation. However, the HRI observation 
was performed in five intervals unevenly spread over 49 hours causing
a large number of alias periods spread over $\sim$0.5 s around the 22.7 s
found from the PSPC observation (which was uninterrupted) and does not 
provide a better estimate for the period.

\subsection{Optical}

We observed the region of \rxjf\ with the ESO-NTT and SUSI2 on 1999 
January 15. The combination of the two EEV CCDs yields a field of view 
of 5\farcm5 $\times$ 5\farcm5 with a pixel size of 0\farcs08. A total of 
30 min in B and 30 min in R were accumulated using 10 min long 
individual exposures. The night was of photometric quality and 
observations of the standard field PG0942-029 allowed photometric 
calibration to an accuracy of $\sim$ 2\% . Mean seeing was 1\arcsec\ 
FWHM. Raw images were corrected for flat-field, bias and cleaned from 
cosmic-ray impacts using standard MIDAS procedures.  

\begin{figure}[]
\psfig{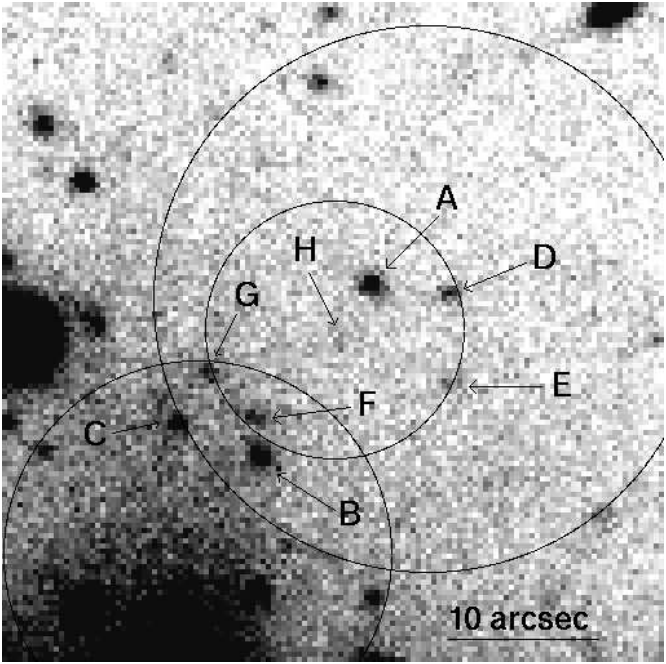} 

\caption[]{The summed B and R images of the field of \rxjf\ corresponding  to a
total integration time of 1 hour and binned with a pixel size of  0.32\arcsec.
The HRI, Survey and PSPC pointing 90\% confidence error circle (8\arcsec ,
12\arcsec and 17\arcsec radius respectively) are over-plotted. Positions of
objects listed in Table 2 are also shown. North is to the top and East to the
left}

\label{'nttfc'} \end{figure}

\begin{figure}[]
\psfig{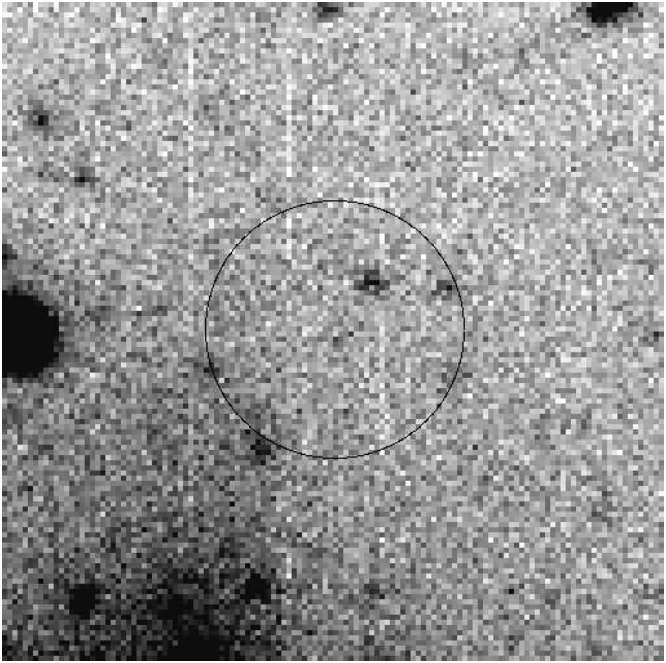} 

\caption[]{The summed B images of the field of \rxjf\ corresponding  to a total
integration time of 30 min and binned with a pixel size of  0.32\arcsec. The
HRI 90\% confidence error circle (8\arcsec\ radius) is over-plotted. North is
to the top and East to the left}

\label{'nttfcb'} \end{figure}

\begin{figure}[]
\psfig{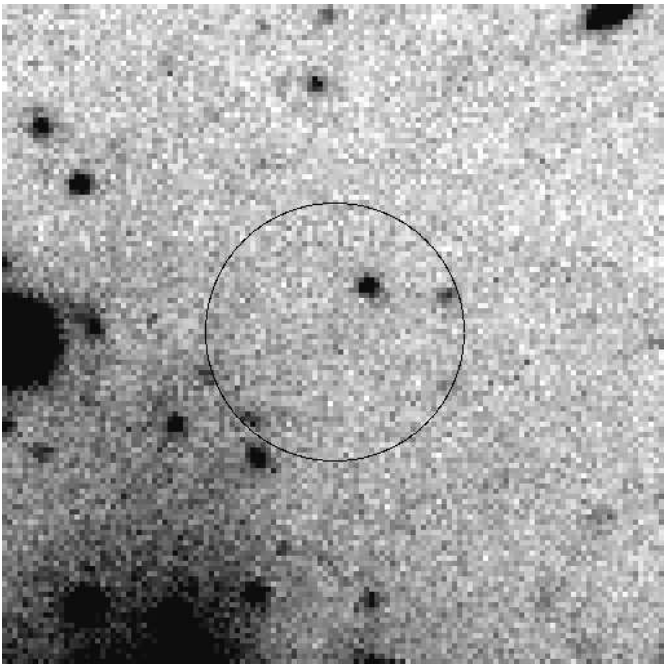} 

\caption[]{The summed R images of the field of \rxjf\ corresponding  to a total
integration time of 30 min and binned with a pixel size of  0.32\arcsec. The
HRI 90\% confidence error circle (8\arcsec\ radius) is over-plotted. North is
to the top and East to the left}

\label{'nttfcr'} \end{figure}

Fig.~\ref{'nttfc'} shows the ROSAT HRI, PSPC and Survey error circles 
overlayed on the sum of the B and R images of \rxjf\ representing a total
integration time  of 1 hour. We also show in Fig.~\ref{'nttfcb'} and
Fig.~\ref{'nttfcr'} the individual B and R summed images with the HRI error
circle overlayed. Optical positions were computed using 8 stars extracted from
the USNO A-1.0 catalogue and spread around the X-ray positions. The final
astrometric accuracy should be that of the USNO catalogue, i.e. better than 1
arcsec.   

Several faint objects labeled A to H on Fig.~\ref{'nttfc'} are detected in or
close to the HRI error circle. Table ~\ref{'om'} lists the B and R magnitudes
with one $\sigma$ errors and information on image profile when possible.
                             
The brightest object in the HRI error circle (object A) is stellar like and
has a B-R colour index suggesting a remote G5V-K0V late type star. The next two
brightest objects (B and D) located at the edge of the error circle appear
clearly extended and are probable field galaxies. Among the remaining
candidates, the brightest one which lacks reliable B-R colour index and
spatial extent is object E with B =  25.25 $\pm$ 0.20. Our detection limit
estimated from magnitudes of object H is B $\sim$ 26.5 and R $\sim$ 26.4.  

\section{Discussion}
The soft X-ray source \rxjf\ is characterized by the same properties as the
five previously published isolated neutron star candidates discovered in
ROSAT data. A blackbody-like spectrum with kT = 57 eV and little interstellar
absorption of 2\hcm{20} are well within the range observed from the known
candidates. No significant long-term flux variations were observed between
ROSAT survey (1991, January 23-25) and three pointed observations in 1997. 

Optical NTT imaging failed to identify the counterpart to the X-ray 
source. The brightest stellar-like object in the error circle has B = 
24.44 $\pm$ 0.08 implying log(\fx/\fo) $\geq$ 3.0 (using the observed X-ray 
flux of 6.9\ergcm{-13} and \fo\ = 10$^{-0.4({\rm B}+13.42)}$ from Maccacaro 
\et\ 1988). It exhibits a red B-R 
colour index, compatible with that of a remote late type star and can 
be ruled out on this basis. The next possible candidate has B = 
25.25 $\pm$ 0.20 yielding a log(\fx/\fo) ratio of 3.3. This virtually excludes 
all possible kinds of soft X-ray emitters other than isolated neutron 
stars. In particular an identification with a hot white dwarf or with a 
very soft Seyfert-1 AGN (e.g. Grupe et al. 1998) seems 
to be excluded as this would imply the presence of an optically bright object 
in the error circle. A very distant white dwarf is also incompatible with the 
low observed absorption.

With 0.11 \cts\ detected in the PSPC \rxjf\ is the faintest of the known INS
candidates so far (see Schwope \et\ 1999). It is a factor of 33 fainter 
than \rxja\ which shows a similar blackbody temperature. Scaling the optical 
B magnitude of \rxja\ by this factor yields an expected value of B 
$\sim$ 29.6 for \rxjf, beyond the reach of our imaging. The serendipitous discovery
near an unrelated object is the first of this kind and more INS may
be unrecognized because of positional mis-identification. It is remarkable
that it shows the same soft spectral properties as the previously found INS
candidates like \rxjc\ and \rxje\ which were discovered from dedicated
searches of objects with such properties. This suggests that most INS have
temperatures kT below $\sim120$ eV and that the distribution is not
strongly biased by selection effects.

\begin{table}
\caption{Optical magnitudes}
\label{'om'}
\begin{tabular}{ccccl}
Object  & B           & R              & B-R           &Profile\\ \hline
A   & 24.44$\pm$0.08  & 23.29$\pm$0.06 & 1.15$\pm$0.10 & Stellar \\
B   & 24.54$\pm$0.10  & 23.52$\pm$0.06 & 1.02$\pm$0.12 & Extended \\
C   &  -              & 23.68$\pm$0.07 & -             & Stellar \\
D   & 25.03$\pm$0.20  & 24.32$\pm$0.10 & 0.71$\pm$0.22 & Extended \\
E   & 25.25$\pm$0.20  & 25.20$\pm$0.20 & 0.05$\pm$0.28 & - \\
F   &  -              & 24.43$\pm$0.10 & -             & - \\
G   & 25.39$\pm$0.20  & 24.58$\pm$0.12 & 0.81$\pm$0.23 & - \\
H   & 26.50$\pm$0.50  & 26.40$\pm$0.50 & 0.10$\pm$0.71 & - \\
\hline
\end{tabular}
\end{table}

In the ROSAT PSPC observation of \rxjf\ we detected X-ray pulsations with 
a significance of 4$\sigma$. If this is confirmed \rxjf\ would 
share this remarkable feature with \rxjb. For
none of the four other candidates periodic flux variations were found
although it must be noted that small amplitude variations could not be
detected in the available data due to insufficient statistics.
The pulse period of 22.7 s for \rxjf\ is even longer than that for 
\rxjb\ (8.391 s) and the modulation is deeper ($>$29\% compared to 12$\pm$2\%).

As in the case of \rxjb\ similar conclusions can be drawn for \rxjf, based
on the different scenarios conceivable for its evolution.
If the X-rays are powered by accretion of interstellar matter then the 
magnetic field strength of \rxjf\ should be lower than 5.3\tento{9} 
d$_{100}$ G where d$_{100}$ is the distance in units of 100 pc. This is 
a similarly low limit as obtained for \rxjb\ for a distance of a few hundred 
pc and would indicate magnetic field decay unless the neutron star was 
born with a relatively long spin period (Wang 1997). The lower temperature 
derived for \rxjf\ may indicate a factor of 
$\sim$4 lower accretion rate than for \rxjb. This in turn increases slightly 
the allowed velocity of the neutron star relative to the ambient 
interstellar medium. However as in the case of \rxjb\ the relative 
velocity would probably be a few tens of km s$^{-1}$, well below 100 km 
s$^{-1}$.

In the magnetar model (Heyl \& Hernquist 1998, Kulkarni \& van Kerkwijk 
1998) the factor 2.7 longer pulse period of \rxjf\ compared to \rxjb\ 
implies a different age and/or magnetic field strength of the neutron stars.
Assuming that the neutron stars have spun down by magnetic dipole 
radiation the age is proportional to P$^2$/B$^2$.
A similar magnetic field strength B would imply that \rxjf\ is a factor of
7.3 older than \rxjb\ while for similar age the magnetic field of \rxjf\
is stronger than that of \rxjb\ by the factor 2.7.
A measure of the pulse period derivative \rxjb\ and \rxjf\ will be essential 
to determine the evolutionary status of these isolated neutron stars.

\acknowledgements
The ROSAT project is supported by the German Bundesministerium f\"ur
Bildung und Forschung (BMBF/DLR) and the Max-Planck-Gesellschaft. 

\end{document}